\documentclass[showpacs,amsmath,amssymb,aps,showkeys,floatfix,prd,a4paper]{revtex4}

\usepackage{graphicx}
\usepackage{dcolumn}
\usepackage{bm}
\usepackage{amsfonts}
\usepackage{amssymb,amscd}
\usepackage{color}

\providecommand{\analysisroot}{{\sc ROOT}}

\def\lsim{\raise0.3ex\hbox{$<$\kern-0.75em\raise-1.1ex\hbox{$\sim$}}}
\def\gsim{\raise0.3ex\hbox{$>$\kern-0.75em\raise-1.1ex\hbox{$\sim$}}}

\def\pom{{I\!\!P}}

\def\beq{\begin{equation}}
\def\eeq{\end{equation}}
\def\bea{\begin{eqnarray}}
\def\eea{\end{eqnarray}}
\def\bq{\begin{quote}}
\def\eq{\end{quote}}

\providecommand{\heppdt}{{\sc HepPDT}}

\providecommand{\hepmc}{{\sc HepMC2}}

\def\gappeq{\mathrel{\rlap {\raise.5ex\hbox{$>$}}
{\lower.5ex\hbox{$\sim$}}}}

\def\lappeq{\mathrel{\rlap{\raise.5ex\hbox{$<$}}
{\lower.5ex\hbox{$\sim$}}}}

\def\Toprel#1\over#2{\mathrel{\mathop{#2}\limits^{#1}}}

\def\pom{{I\!\!P}}

\begin{document}

\title{Exclusive dilepton production at forward rapidities in $PbPb$ collisions}

\author{V. P. Goncalves$^{1}$, D. E. Martins$^2$,  M. S. Rangel$^{2}$}

\affiliation{$^{1}$ Instituto de F\'{\i}sica e Matem\'atica,  Universidade
Federal de Pelotas (UFPel), \\
Caixa Postal 354, CEP 96010-900, Pelotas, RS, Brazil}

\affiliation{$^{2}$ Instituto de F\'isica, Universidade Federal do Rio de Janeiro (UFRJ), 
Caixa Postal 68528, CEP 21941-972, Rio de Janeiro, RJ, Brazil}


\begin{abstract}
The dilepton production in diffractive and exclusive processes at forward rapidities considering ultraperipheral $PbPb$ collisions at the LHC is investigated. Predictions for the $e^+ e^-$, $\mu^+ \mu^-$ and $\tau^+ \tau^-$ cross sections are presented taking into account of realistic cuts that can be implemented by the LHCb Collaboration in a future experimental analysis. Our results indicate that the background associated with the diffractive production can be strongly suppressed and the exclusive processes can be cleanly separated. For the $\tau^+ \tau^-$ production, the semi and purely leptonic decay channels are considered. Our results indicate that a future experimental analysis of the dilepton production at the LHCb is feasible and can be useful to  search for BSM physics.
\end{abstract}


\maketitle

The experimental study of photon -- induced interactions in $pp/pA/AA$ collisions became a reality in the last years,  motivated by the possibility of improve our understanding of the Standard Model as well as by the opportunity of use these processes as an alternative tool to search for New Physics (for a recent review see e.g. Ref. \cite{klein}). 
Recent results for the exclusive diphoton production in ultraperipheral heavy ion collisions (UPHICs) \cite{upc}  provided the first observation of the  light -- by -- light (LbL) scattering \cite{Aad:2019ock,Sirunyan:2018fhl}, which was one of the most important predictions in the beginning of Quantum Electrodynamics (QED). Another important result was the observation of the exclusive dilepton production by photon -- photon  interactions in
heavy ion collisions at the RHIC \cite{star_dilepton} and LHC \cite{alice_dilepton,atlas_dilepton, cms_dilepton}, which has demonstrated the feasibility of study this final state and allowed 
 us to improve the description of the nuclear photon flux and the modeling of the absorption effects, that suppress the strong interactions in these collisions \cite{mariola,celsina}. Moreover, both results strongly motivated the studies related to beyond Standard Model (BSM) physics  performed in Refs.~\cite{knapen,royon,nosplb,antonitau,liu}, which focused to estimate the impact of the axionlike particle production in the diphoton final state and in the possibility of determine the $\tau$ lepton electromagnetic moments in the $\gamma \gamma \rightarrow \tau^+ \tau^-$ interactions present in $PbPb$ collisions at the LHC. In particular, the results obtained in Refs.~\cite{antonitau,liu} have demonstrated the potentiality of the exclusive dilepton production to constrain BSM physics.

The studies performed in \cite{mariola,celsina,antonitau,liu} were restricted to the rapidity range 
covered by the central detectors as e.g. ALICE, ATLAS and CMS. Such selection is justified, since the rapidity distribution is larger for midrapidities. However, for the LHC energies, the exclusive dilepton production for forward rapidities can become of the same order, especially if  dileptons with small values of invariant masses and low transverse momenta are measured, as possible in the LHCb detector.  	
One of the goals of this letter is to present, for the first time, predictions for the exclusive $e^- e^+$, $\mu^- \mu^+$ and $\tau^- \tau^+$ production at forward rapidities considering realistic selection cuts that can implemented by the LHCb Collaboration to select the associated events. As we will show below, a future experimental analysis is feasible. Our second goal is to estimate, for the first time, the background associated to dilepton production in double diffractive processes in $PbPb$, which also are characterized by two intact nuclei and two rapidity gaps in the final state,  i.e., empty regions  in pseudo-rapidity that separate the intact very forward nuclei from the $l^+ l^-$ state. We will demonstrate that such background can be strongly suppressed by the exclusivity cuts, which implies that future data will be a direct probe of the dilepton production by photon -- photon interactions. As a by product, we also will show that the diffractive events can be separated using a different set of cuts (For a similar analysis for the diphoton production see Ref. \cite{nosdiphoton}).

\begin{figure}[t]
\begin{center}
\begin{tabular}{ccc}
\includegraphics[scale=0.43]{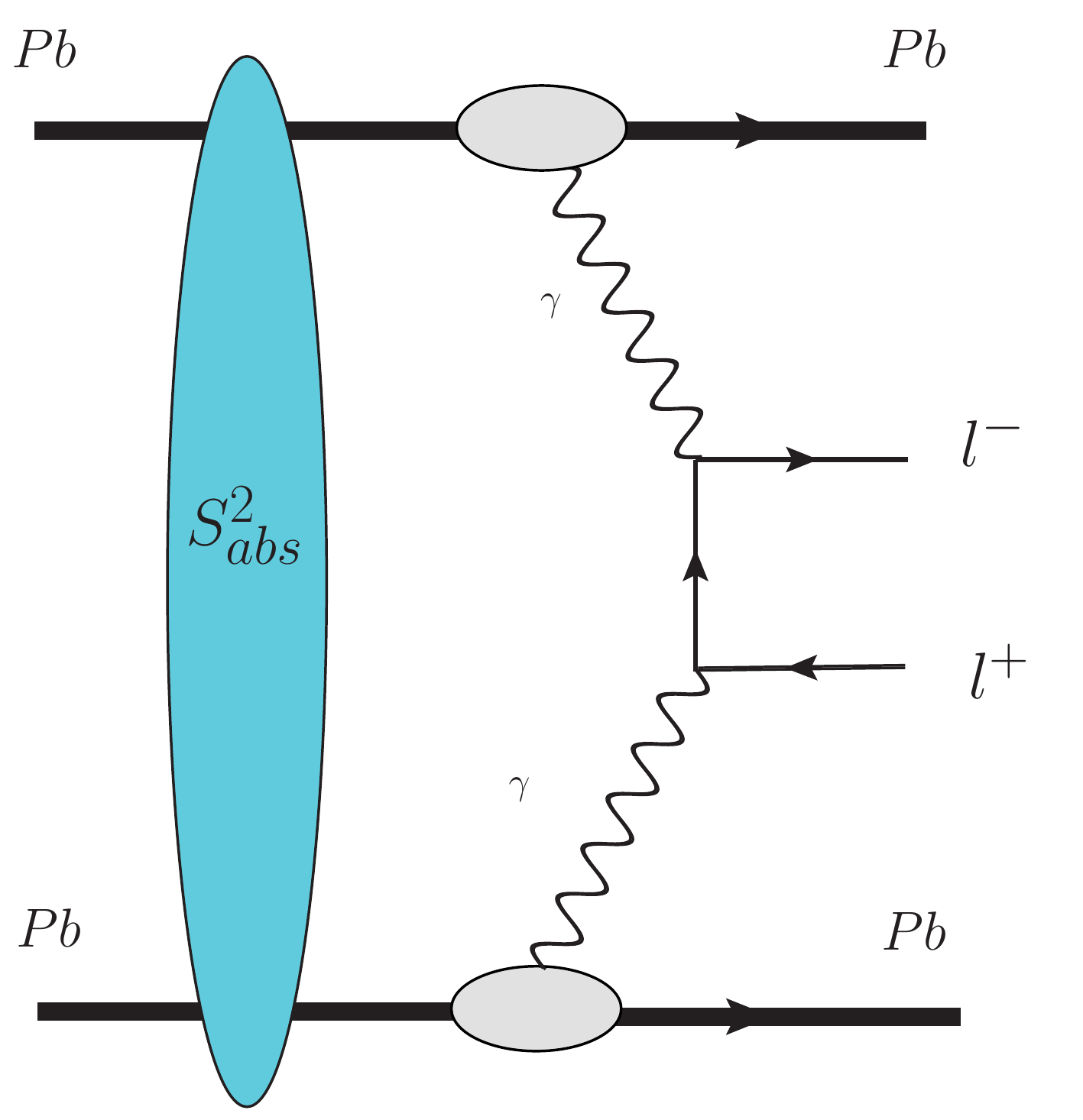} & \,\,\, &
\includegraphics[scale=0.34]{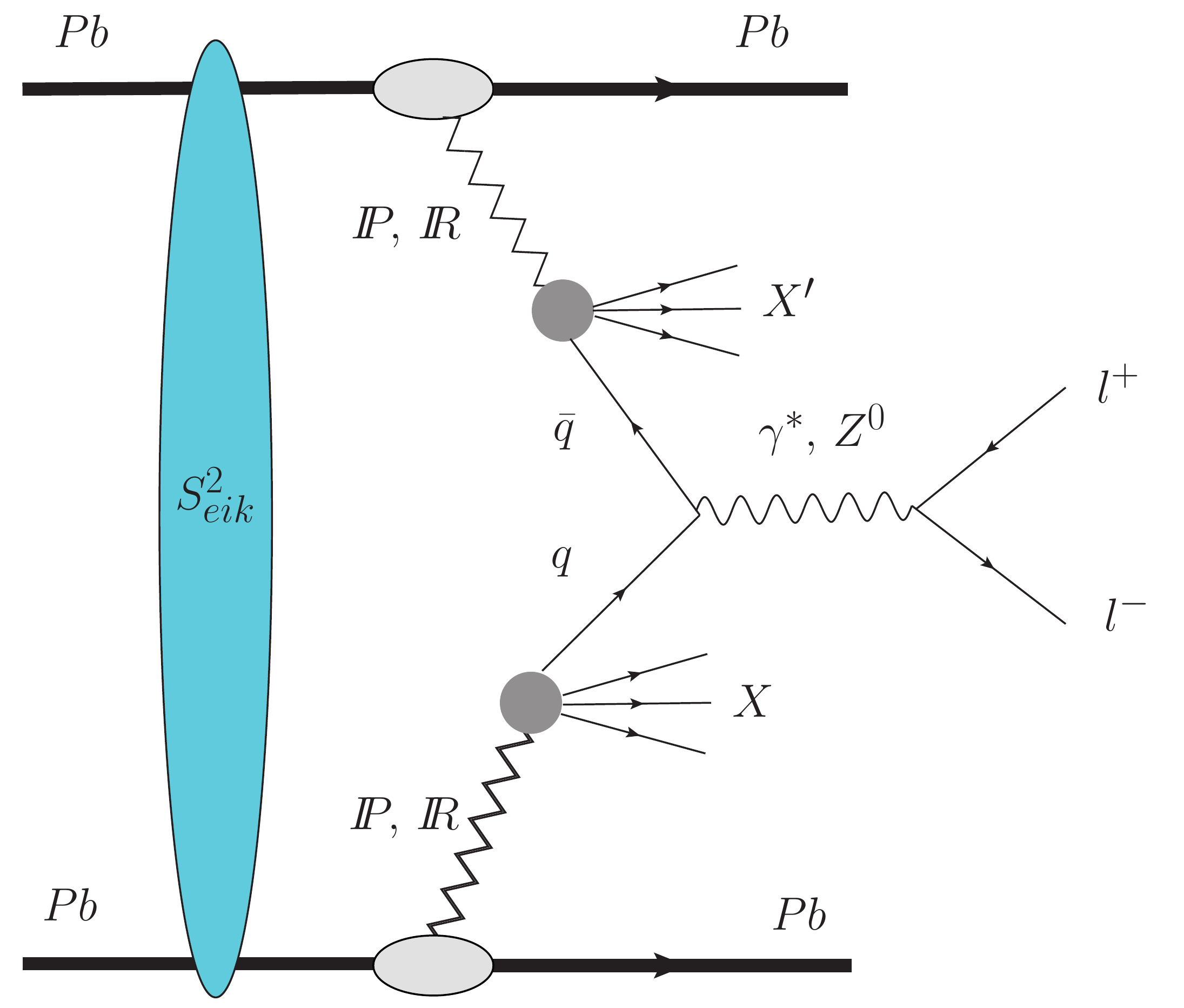} \\
(a) & \,\,\, &  (b)
\end{tabular}
\caption{Dilepton production in (a) exclusive and (b) diffractive interactions considering $PbPb$ collisions at the LHC.
}
\label{Fig:diagram}
\end{center}
\end{figure}

In what follows we will present our predictions for the exclusive dilepton production by photon -- photon  interactions in ultraperipheral heavy ion collisions, represented in Fig. \ref{Fig:diagram} (a), which are characterized by an impact parameter $b$ greater than the sum of the radius of the incident nuclei. In such collisions, the strong interactions between the incident nucleus are suppressed and the  electromagnetic interaction between them become dominant  \cite{upc,upc_dilepton}.
The cross section for the $PbPb \rightarrow Pb \, l^+ l^- \,Pb $  can be estimated using the equivalent photon approximation \cite{epa}, which allow us to factorize the elementary $\gamma \gamma \rightarrow l^+ l^-$ process and  describe the intense electromagnetic fields that accompany the relativistic heavy ions in terms of  a spectrum of equivalent photons proportional to the square of the nuclear charge $Z$. As a consequence, the nuclear cross section is proportional  to $Z^4$, implying a very large number of events at RHIC and LHC. Moreover, it is assumed  that the photons  are coherently radiated by the whole nucleus. Such condition imposes that the minimum photon wavelength must be greater than the nuclear radius $R$. In other words, the photon virtuality must satisfy $Q^2 = -q^2 \le 1/R^2$, with the photon four -- momentum being $q^{\mu} = (\omega, \vec{q_{\perp}},q_z = \omega/v)$, where $\vec{q_{\perp}}$ is the transverse momentum of the photon in a given frame, where the projectile moves with velocity  $v$. Therefore, we have that $Q^2 = \omega^2/\gamma_L^2 + q_{\perp}^2$.
The coherence condition  limits the maximum energy of the photon to $\omega < \omega_{\mbox{max}}  \approx \gamma_L/R$ and the perpendicular component of its momentum to ${q_{\perp}} \le 1/R$. Therefore, the coherence condition  sets an upper limit on the transverse momentum of the photon emitted by the hadron, which should satisfy $q_{\perp} \le 1/R$, being $\approx 28 $ MeV for $Pb$ beams. One has that the photon virtuality can be neglected and the photons can be considered as being real. Moreover, the dilepton pair will be produced with a very small transverse momentum. 
In the impact parameter representation, the total cross section can be expressed as follows
\begin{eqnarray}
\sigma \left(Pb Pb \rightarrow Pb \otimes l^+ l^- \otimes Pb;s \right)   
&=& \int \mbox{d}^{2} {\mathbf r_{1}}
\mbox{d}^{2} {\mathbf r_{2}} 
\mbox{d}W 
\mbox{d}Y \frac{W}{2} \, \hat{\sigma}\left(\gamma \gamma \rightarrow l^+ l^-; 
W \right )  N\left(\omega_{1},{\mathbf r_{1}}  \right )
 N\left(\omega_{2},{\mathbf r_{2}}  \right ) S^2_{abs}({\mathbf b})  
  \,,
\label{cross-sec-2}
\end{eqnarray}
where $\sqrt{s}$ is center - of - mass energy of the $PbPb$ collision, $\otimes$ characterizes a rapidity gap in the final state, 
$W = \sqrt{4 \omega_1 \omega_2}$ is the invariant mass of the $l^+ l^-$ system and $Y$ its rapidity. The photon energies $\omega_1$ and $\omega_2$  are related to   $W$ and to the rapidity  $Y$ of the outgoing dilepton system by $
\omega_1 = \frac{W}{2} e^Y$ and $\omega_2 = \frac{W}{2} e^{-Y}$. The cross section $\hat{\sigma}$ is the elementary cross section to produce a pair of leptons, where l = $e$, $\mu$, $\tau$. Moreover, $N(\omega_i, {\mathbf r}_i)$ is the equivalent photon spectrum   with energy $\omega_i$ at a transverse distance ${\mathbf r}_i$ from the center of nucleus, defined in the plane transverse to the trajectory, which is determined by the charge form factor of the nucleus. Finally,  the factor $S^2_{abs}({\mathbf b})$ depends on the impact parameter ${\mathbf b}$ of the $PbPb$ collision and  is denoted by absorptive factor, which excludes the overlap between the colliding nuclei and allows to take into account only ultraperipheral collisions. As in Refs.\cite{celsina,nosdiphoton}, the nuclear photon flux will be estimated assuming a point -- like form factor and that the absorptive factor $S^2_{abs}({\mathbf b})$ will be described by the model proposed by Baur and Ferreira - Filho \cite{Baur_Ferreira} to exclude the strong interactions between the incident nuclei. One has that the predictions for small invariant masses, which is the kinematical range of interest in this letter, are not dependent on these choices, as demonstrated in Ref.\cite{celsina}. In  our calculations we express Eq.(\ref{cross-sec-2}) in the momentum space in order to impose cuts that depend on the lepton transverse momenta. Moreover, our analysis will be performed using  the Forward Physics Monte Carlo (FPMC) \cite{fpmc}, which allow us to estimate the associated cross sections and distributions taking into account of the detector acceptances.

One possible background for the exclusive dilepton production in UPHICs is the Drell - Yan (DY)  production in double diffractive processes, represented in Fig.~\ref{Fig:diagram} (b). As shown in Ref.~\cite{nosdilepton_pp} for $pp$ collisions, such process can dominate in some regions of the phase space. However, its contribution for heavy ion collisions is still an open question. In the diffractive DY process, the dilepton is  produced through the $q \bar{q} \rightarrow l^+ l^-$ subprocess, with the partons being constituents of the Pomerons emitted by the incident nuclei, which remain intact after the emission. As the Pomeron is a color -- singlet object, two rapidity gaps are expected in the final state, as in the case of the exclusive dilepton production. The main difference between these processes is the presence of the Pomeron remnants $X$ and $X^{\prime}$ in the double diffractive case, which are expected to generate additional tracks in the final state, in contrast with the exclusive case, where only the dilepton state is present. Moreover, from the analysis of the diphoton production performed in Ref. \cite{nosdiphoton}, we can expect that the transverse momentum distribution of the dileptons in these two channels will be  different, with the DY process generating dileptons with larger transverse momentum. Such aspects will be considered in what follows in order to suppress the contribution of the diffractive process in comparison to the exclusive one. As in Ref. \cite{nosdiphoton}, we will estimate the cross section for the diffractive DY production using the Resolved Pomeron  model \cite{IS}, which implies that the cross section can be expressed by
\begin{eqnarray}
\sigma(Pb Pb \rightarrow Pb \otimes X+l^+ l^-+X^{\prime} \otimes Pb)  =   S^2_{eik} & \times & \int dx_{1} \int  dx_{2} \, 
\, \left[q^D_{1}(x_{1},\mu^2) \cdot \bar{q}^D_{2}(x_{2},\mu^2) \right. \nonumber \\
 & + & \left. \bar{q}^D_{1}(x_{1},\mu^2) \cdot  {q}^D_{2}(x_{2},\mu^2) \right] \cdot \hat{\sigma}(q \bar{q} \rightarrow l^+l^-) ,
\label{pompom}
\end{eqnarray}
where $ S^2_{eik}$ is the soft survival factor, which takes into account of the soft interactions which lead to an extra production of particles that destroy the rapidity gaps in the final state \cite{bjorken}, and  $q^D_i (x_i,\mu^2)$ and $\bar{q}^D_i (x_i,\mu^2)$ are the diffractive quark and antiquark densities of the nucleus $i$ with a momentum fraction $x_i$. Such quantities will be modeled using the approach proposed in 
\cite{vadim} (See also Ref. \cite{review_vadim}), which have described the diffractive distributions for a nucleus   taking into account the nuclear effects associated to the nuclear coherence and the leading twist nuclear shadowing (For details see Ref. \cite{nosdiphoton}). In our analysis we will estimate the associated cross section and distributions using the FPMC generator, where the hard matrix elements are treated by interfacing FPMC with HERWIG  v6.5.
Moreover, we will assume that $S^2_{eik} = 3.4 \times 10^{-5}$ for $PbPb$ collisions at $\sqrt{s} = 5.5$ TeV, which was calculated in Ref. \cite{nos_dijet}, using an approach that  generalizes the model described in Ref. \cite{berndt} for coherent double exchange processes in nuclear collisions.

 \begin{center}
 \begin{figure}[t]
 \includegraphics[width=0.31\textwidth]{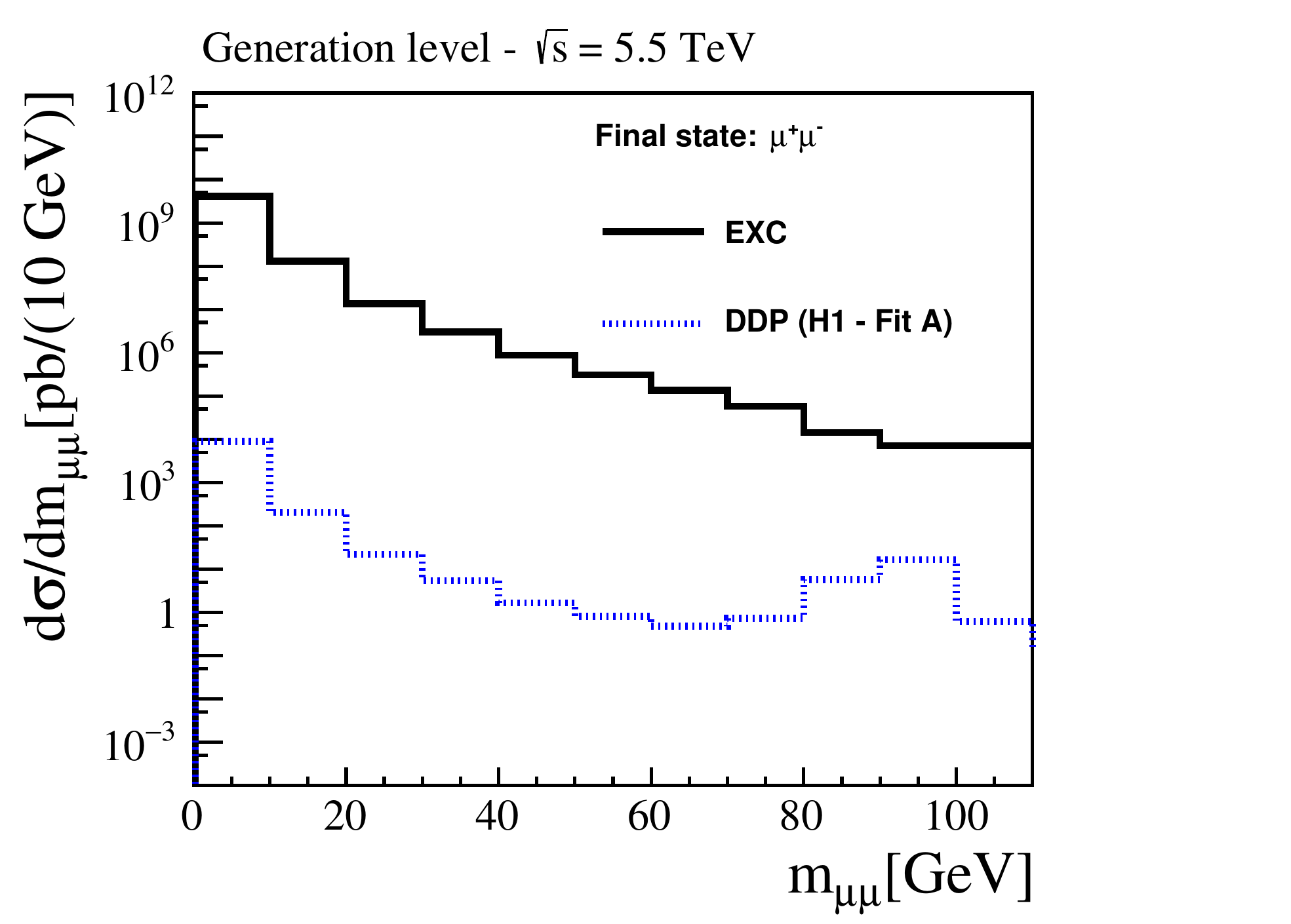}
 \includegraphics[width=0.31\textwidth]{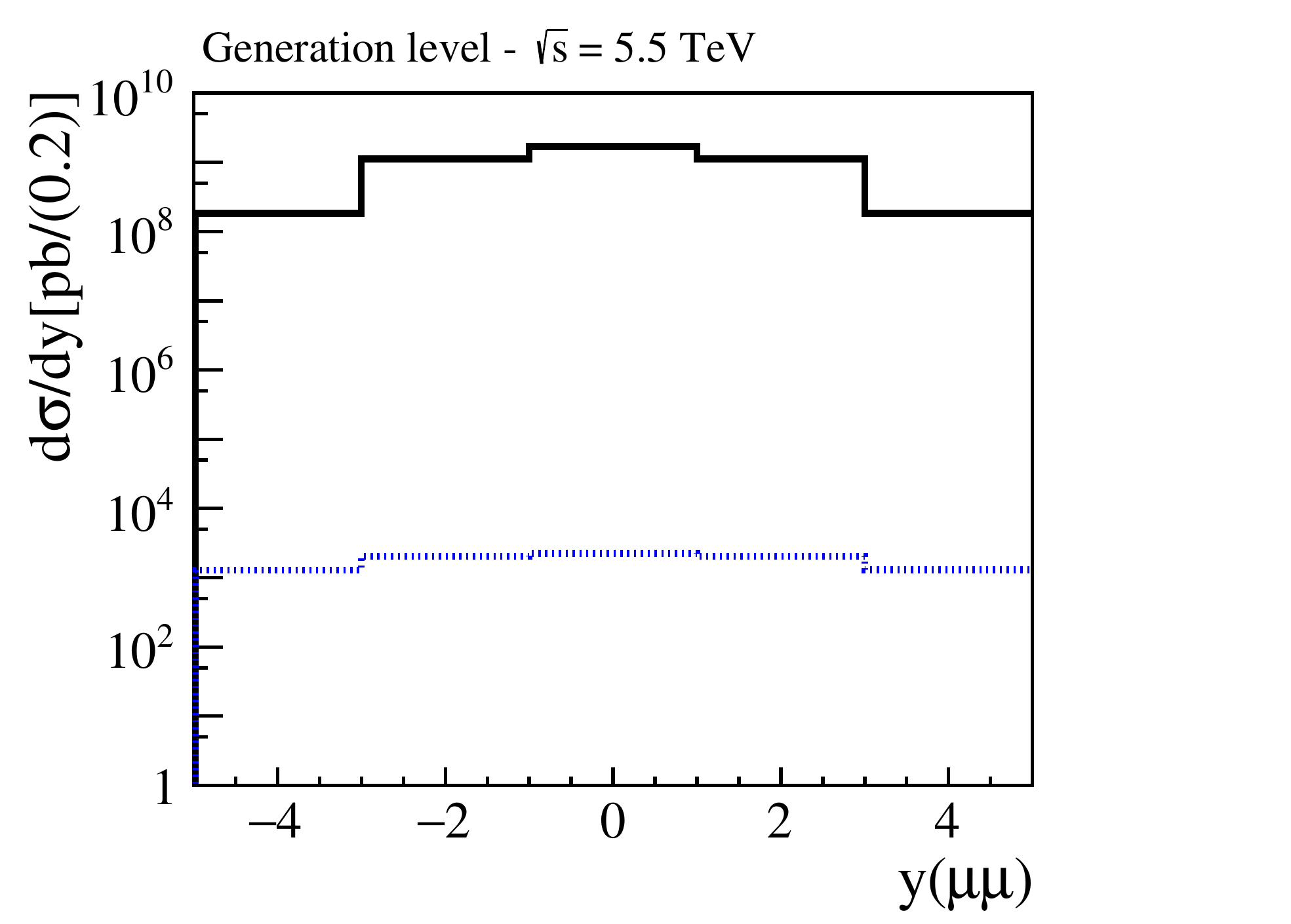}
 \includegraphics[width=0.31\textwidth]{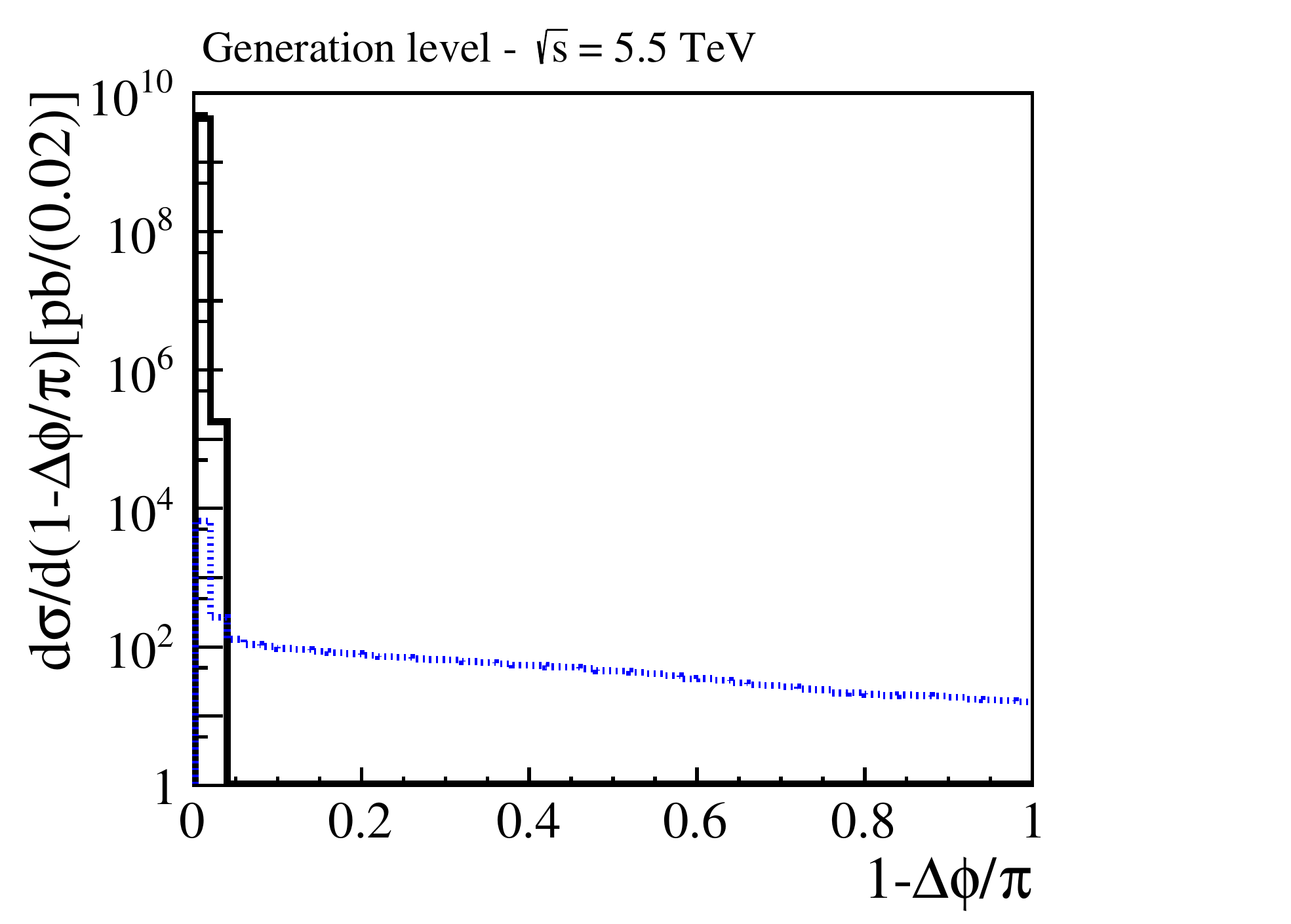}
 \caption{Results for the invariant mass $m_{ll}$, transverse momentum $p_T (pair)$, acoplanarity and rapidity $y(ll)$ distributions of dileptons $\mu^{+}\mu^{-}$ at generation level for $\sqrt{s}=5.5$ TeV in $PbPb$ collisions at the LHC.}
\label{fig:genlevel}
 \end{figure}
 \end{center}

In what follows we will present our results for the exclusive and diffractive dilepton production in $PbPb$ collisions at $\sqrt{s} = 5.5$ TeV. In our analysis we will use the FPMC event generator \cite{fpmc} to estimate the processes represented in the Figs. \ref{Fig:diagram} (a) and (b).
 For the final state selection we use \hepmc~\cite{hepmc} with a plugin called \heppdt \cite{heppdt} which has been designed to be used by any Monte Carlo particle generator or decay package. \heppdt~ has the function to store particle information such as charge and nominal mass in a table which is
accessed by a particle ID number. The particle ID number is defined according to the Particle Data Group’s Monte Carlo numbering scheme~\cite{pdg}.
The final analysis and distributions are done with \analysisroot~\cite{root}. We will focus on  dileptons with  invariant mass in the range $1.5\, \mbox{GeV} \le  m_{ll} \le 20$ GeV, removing the dileptons with invariant mass in the ranges $2.796\, \mbox{GeV} \le m_{ll} \le 3.196\, \mbox{GeV}$,  $3.586\, \mbox{GeV} \le m_{ll} \le 3.786\, \mbox{GeV}$, and   
$9.0\, \mbox{GeV} \le m_{ll} \le 10.6\, \mbox{GeV}$, associated to  the $J/\Psi$, $\Psi(2S)$ and $\Upsilon$ decays, respectively.

Initially, let's consider the dimuon production and  present in Fig. \ref{fig:genlevel}  our results for the  invariant mass ($m_{\mu \mu}$),  rapidity ($y_{\mu \mu}$) and acoplanarity ($1 -(\Delta \phi/\pi)$) distributions considering the exclusive and diffractive  channels, which were obtained at the generation level, without the inclusion of any selection in the events. We have that the diffractive contribution is strongly suppressed by the soft survival effects, with the associated invariant mass and  rapidity  distributions  being a factor $\approx 10^5$ smaller than the photon -- induced one in the $m_{\mu \mu}$ and $y_{\mu \mu}$ ranges considered. On the other hand, the results for the acoplanarity distribution indicate that in  the exclusive case the muons are dominantly produced in a configuration back - to - back. Such result is expected due to the very small transverse momentum carried by the photons in the initial state. In contrast, for the diffractive case, the transverse momentum of the quarks inside the Pomeron that interact to generate the dimuon can be large, which implies  a non - negligible contribution associated to muon pairs with a large acoplanarity. Similar results are obtained for the $e^+ e^-$ and $\tau^+ \tau^-$ production.

In order to obtain realistic estimates of the dilepton production in $PbPb$ collisions, which can be compared with the future experimental data, we will include in our analysis the experimental cuts  that are expected to be feasible in the next run of the LHC by the LHCb detector. 
Due to the distinct characteristics of the $\tau$ decay, we will study this final state using a different strategy to be discussed below. For the 
$e^+ e^-$ and $\mu^+ \mu^-$ production, the selection criteria implemented in our analysis of double diffractive and exclusive diphoton processes at forward rapidities are the following:   
\begin{itemize}
\item The initial selection is made on the transverse momentum of the leptons and invariant mass: $m_{ll} <$ 20 GeV and $p_{T}(l,l) >$ 0.4 GeV. The notation $p_{T}(l,l)$ represents that the transverse momentum cut is applied in both leptons separately;
\item  We will impose a cut on the acoplanarity ($1 -(\Delta \phi/\pi) <$ 0.03) and transverse momentum of the dilepton system ($p_{T}(l l) <$ 2 GeV);
\item We will select only events where leptons are produced in the rapidity range $2.0 < \eta(l,l) < 4.5$ and no other particle with $p_T > 0.5$ GeV in the range $8.0 < |\eta| <  5.5$ is produced (this  corresponds to the HERSCHEL acceptance installed in the LHCb experiment~\cite{herschel}).
\end{itemize}  
The associated predictions are presented in the Table \ref{table:sigma_leves}. One has that the predictions for the dielectron and dimuon production are very similar, which is expected in the invariant mass range considered, and the cut on the transverse momentum of the pair suppress the diffractive contribution. In addition, the HERSCHEL selection also suppresses this contribution and  does not impact the exclusive one. Such results are  expected, since in the exclusive case there are no additional tracks beyond those associated with the dilepton, in contrast to the diffractive production, where the remnants of the Pomeron are expected to generate new tracks. One important comment is that the cut on rapidity for the kinematical range covered by the LHCb detector reduces the cross section by approximately one order of magnitude. However, the final values are still very large, implying that the experimental analysis is feasible.

\begin{center}
\begin{table}[!t]
\begin{tabular}{ l||c|c|c|c|}
\hline 
\hline
 PbPb @ 5.5 TeV & \multicolumn{2}{|c|}{{ $\sigma(e^{+}e^{-})$ [pb]}}  & \multicolumn{2}{|c|}{{$\sigma(\mu^{+}\mu^{-})$ [pb]}} \\
\hline
Process  & Diffractive  & Exclusive & Diffractive& Exclusive \\
\hline
Generation    &  9180.0   & $4.3\times10^{9}$  & 9180.0 & $4.3\times10^{9}$\\
\hline
  1.5 GeV $<m_{ll}^{*} < 20$ GeV, $p_{T}^{l}> 0.4$ GeV &6948.0 &$3.4\times10^{9}$  & 6928.0 &$3.4\times10^{9}$ \\
 \hline 
 $p_{T}(ll)<$ 2 GeV & 2602.0  & $2.8\times10^{7}$ &2610.0 & $3.0\times10^{7}$\\
 \hline 
 Acoplanarity $<$ 0.03 & 2476.0   & $2.8\times10^{7}$  & 2480.0&$3.0\times10^{7}$\\
 \hline 
 $ 2.0 \leq \eta_{l,l} \leq 4.5$ & 187.0   & $2.6\times10^{6}$   &187.0&  $1.8\times10^{6}$ \\
 \hline 
 HERSCHEL  &  11.0  &  $2.6\times10^{6}$    &  12.0   &$1.8\times10^{6}$\\
 \hline 
\end{tabular}
\caption{Predictions for the  dielectron and dimuon production at forward rapidities in $PbPb$ collisions at $\sqrt{s} = 5.5$ TeV  in exclusive and diffractive interactions considering kinematical cuts that select the {\it exclusive} contribution. }
\label{table:sigma_leves}
\end{table}
\end{center}

The results presented in Fig. \ref{fig:genlevel} indicated that the {acoplanarity} distribution for exclusive and diffractive processes are distinct. Therefore we can invert the acoplanarity cut used in our previous analysis in order to select the diffractive events instead of the exclusive one. The results for the dimuon production derived using this new set of cuts are presented in the Table \ref{table:sigma_diff}. Similar results are obtained for the dielectron production. One has that the exclusive contribution is strongly suppressed and the measurement of dileptons for these selection criteria is a direct probe of the diffractive channel. Such result opens the possibility to constrain the magnitude of the soft survival factor $ S^2_{eik}$ in the dilepton production, which is one of the main sources of theoretical uncertainty in the treatment of diffractive processes at nuclear collisions.

\begin{center}
\begin{table}[!t]
\begin{tabular}{ l||c|c|}
\hline 
\hline
 PbPb @ 5.5 TeV &  \multicolumn{2}{|c|}{{$\sigma(\mu^{+}\mu^{-})$ [pb]}} \\
\hline
Process  & Diffractive  & Exclusive   \\
\hline
Generation    & 9180.0   &$4.3\times10^{9}$\\
\hline
  1.5 GeV $<m_{ll}^{*} < 20$ GeV, $p_{T}^{l}> 0.4$ GeV &  6948.0 & $3.4\times10^{9}$ \\
 \hline 
 Acoplanarity $>$ 0.03       &  1829.0&0.0\\
 \hline 
 $ 2.0 \leq \eta_{l,l} \leq 4.5$  &206.0 & 0.0  \\
 \hline 
 \hline 
\end{tabular}
\caption{Predictions for the   dimuon production at forward rapidities in $PbPb$ collisions at $\sqrt{s} = 5.5$ TeV  in exclusive and diffractive interactions considering kinematical cuts that select the {\it diffractive} contribution. }
\label{table:sigma_diff}
\end{table}
\end{center}

In what follows we will discuss the exclusive and diffractive $\tau^+ \tau^-$ production in UPHICs. The exclusive case was recently investigated in Refs. \cite{antonitau,liu}, strongly motivated by the possibility of use this final state to constrain the anomalous magnetic momentum of the $\tau$ lepton and search for CP violation effects. These studies focused on the kinematical range covered by the ATLAS and CMS detectors. We will complement these analyzes, by analyzing the possibility of measure this final state in the LHCb detector, which collected data with low pile-up and it is able to reconstruct low momentum tracks.  We will restrict our study to the Standard Model subprocesses and postpone for a future publication the investigation of the impact of new physics.  Differently from the lighter  charged leptons, the  very short lifetime  of the $\tau$ lepton [${\cal{O}}(10^{-13}$ s)] implies that it decays into lighter leptons  and hadrons before to interact with the detector. As a consequence, the separation of the $\gamma \gamma \rightarrow \tau^+ \tau^-$ events depends on a precise identification of the decay products. The $\tau$ decays can be classified as one prong decays ($\approx 80 \%$), where one charged particle is present in the final state, and three prong decays ($\approx 20 \%$), which are characterized by the production of three charged particles.  
In our analysis, only the 1-prong decays (semi and purely leptonic) production will be considered. Moreover, in order to suppress the contribution of the diffractive DY process for the purely leptonic channel, we will select events where the leptons in the final state have different flavours. 
Therefore, the following events will be included in our study:
\begin{eqnarray}
\tau\tau \rightarrow (e^{\pm}\nu_{e}\nu_{\tau})\,(\mu^{\mp}\nu_{\mu}\nu_{\tau}) ,\,\, (e^{\pm}\nu_{e}\nu_{\tau})\, (\pi^{\mp}\pi^{0}\nu_{\tau}),\:\:  (\mu^{\pm}\nu_{\mu}\nu_{\tau})\, (\pi^{\mp}\pi^{0}\nu_{\tau}) \:.
\end{eqnarray}
The HERWIG 6.5 program \cite{herwig}, which incorporates the TAUOLA MC code \cite{tauola}, is used to model $\tau$ decay for our studies.
We select events with  $p_{T}(l,chg) > 1.0$ GeV and  $2.0 < \eta^{l,chg} < 4.5$ with 0 extra tracks and $p_{T}(l + chg) < 2.0 $  GeV. 
As considered for lighter leptons, we discard events where particles with $p_T > 0.5$ GeV are in the range $ 8.0 < |\eta| <  5.5$, corresponding to the HERSCHEL acceptance in the LHCb experiment~\cite{herschel}. Before presenting our predictions, it is important to emphasize that we also have performed the analysis considering the rapidity range covered for central detectors and obtained a very good agreement with the results presented in Ref. \cite{antonitau}.

 \begin{center}
 \begin{figure}[t]
 \includegraphics[width=0.45\textwidth]{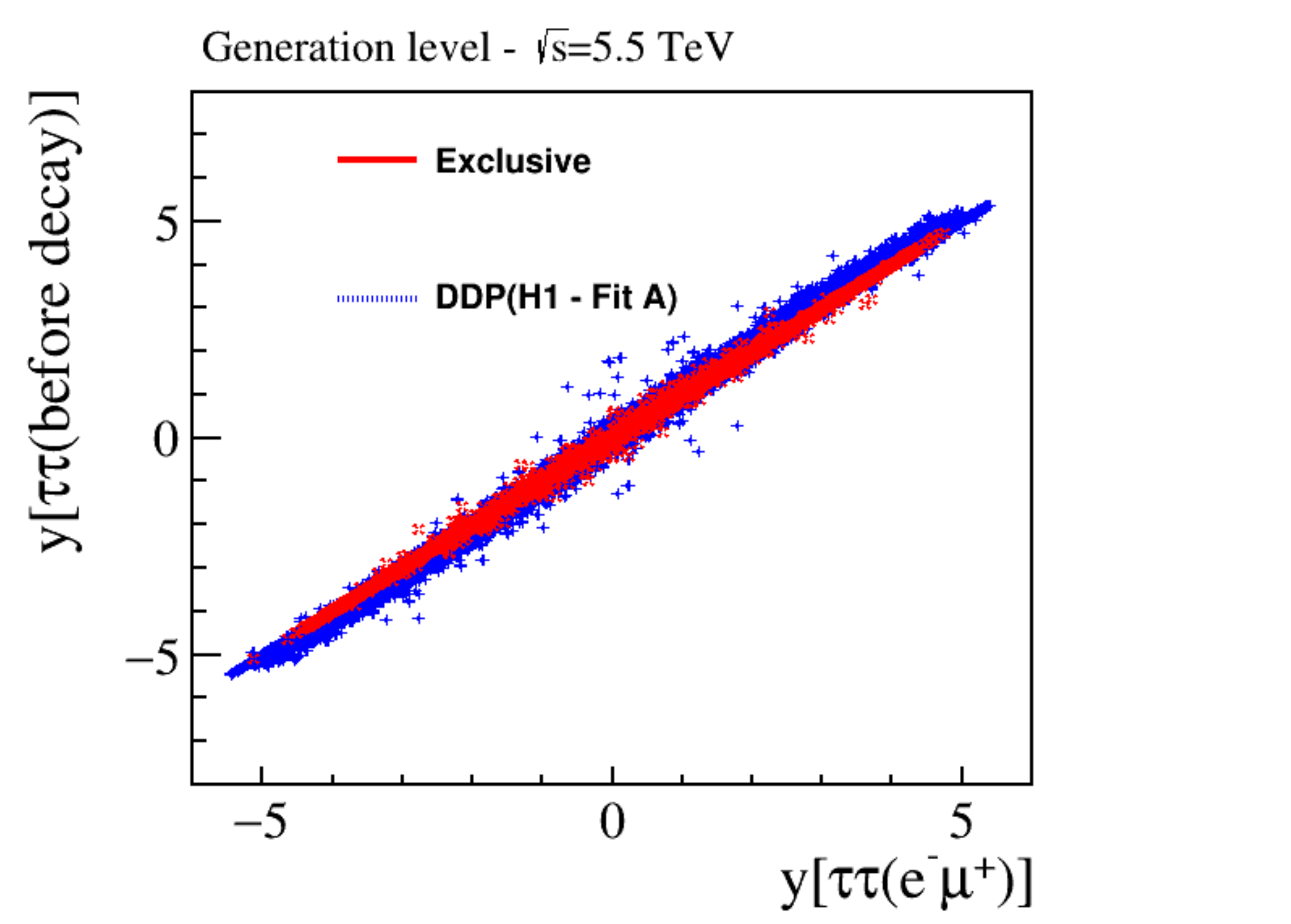}
 \includegraphics[width=0.45\textwidth]{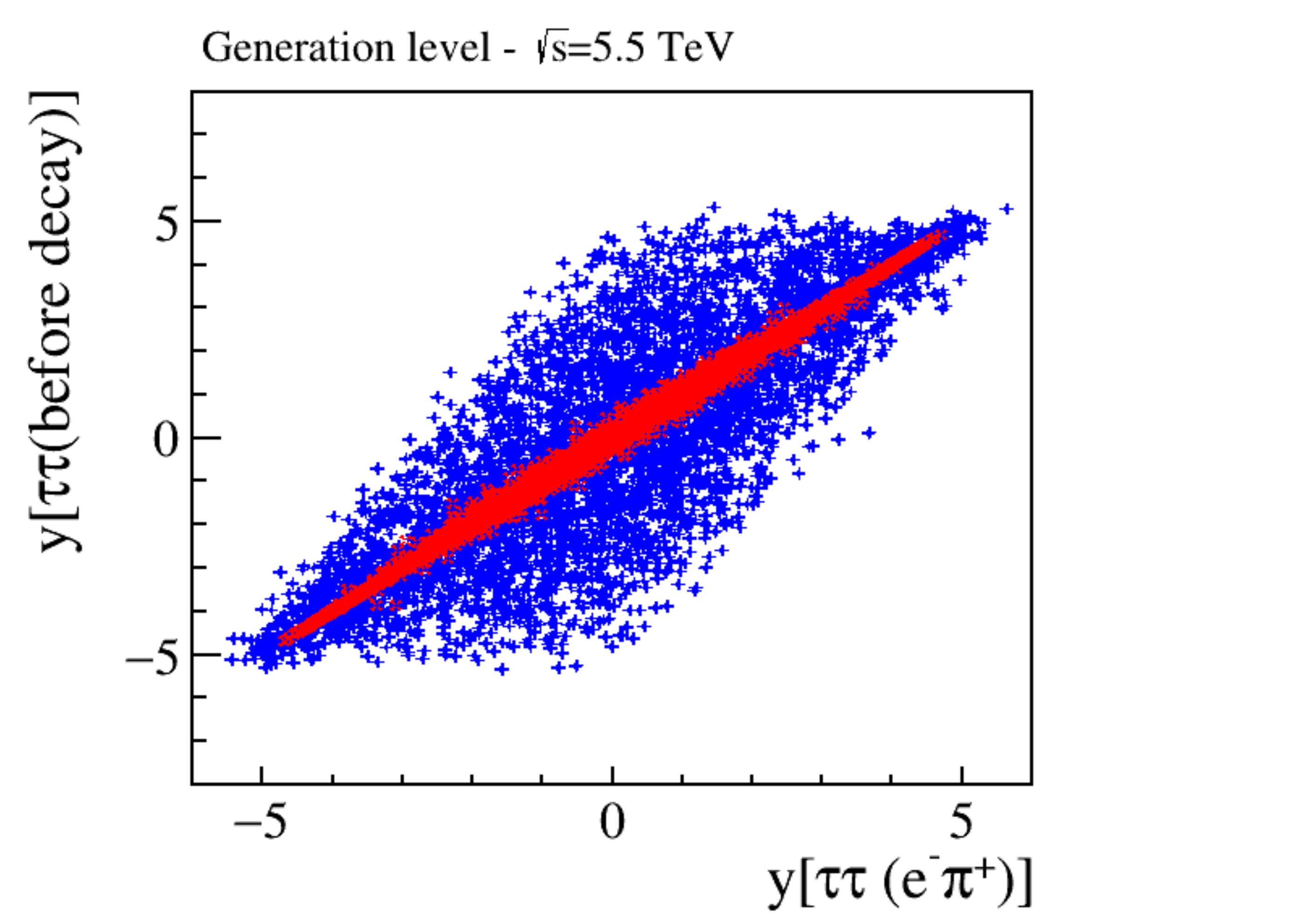}
 \caption{Correlation between the rapidity of the $\tau^+ \tau^-$ system  and the purely leptonic pair (left panel) and the semileptonic system (right panel) generated in the decays for the 1-prong category.  
 }
\label{fig:correlation_tau}
 \end{figure}
 \end{center}

In Fig. \ref{fig:correlation_tau} we present our results for  the correlation between the hard process and reconstruction of the $\tau^{+}\tau^{-}$ pair considering the 1-prong category at generation level. The reconstruction of  rapidity of the pair is done before the exclusivity cuts and establish a benchmark of the effective combinations resulting in the $\tau^{+}\tau^{-}$ pair products. { Our results indicate that the rapidity of the system before decay and the final state charged particle in  exclusive and diffractive processes are correlated, with this correlation being larger in the exclusive case}. In Table \ref{table:sigma_tau} we present our predictions for the $\gamma \gamma \rightarrow \tau^{+}\tau^{-}$ reaction in $PbPb$ collisions at $\sqrt{s} = 5.5$ TeV considering the different decay channels. As already observed for lighter leptons, the exclusive channel dominates the $\tau^{+}\tau^{-}$ production. Moreover, the semileptonic contribution for the exclusive production is a factor $\approx 2.5$ larger than the purely leptonic one.  Our results indicate that the total cross section for the exclusive $\tau^{+}\tau^{-}$ production in the kinematical covered by the LHCb detector will be $\approx$ 60 nb, which implies that the number of events expected  for the next run of the LHC is $\approx$ 600 per year. Although smaller than the prediction for central detectors, this result indicates that a future experimental analysis is feasible.

\begin{center}
\begin{table}[t]
\begin{tabular}{ |c|c|c|c|c|c|c|c|c|c|c|}
\hline
 PbPb @ 5.5 TeV & \multicolumn{10}{|c|}{{$\sigma(\tau^{+}\tau^{-})$ [pb]}}  \\
\hline
Process  & \multicolumn{5}{|c|}{Diffractive} & \multicolumn{5}{|c|}{Exclusive} \\
\hline
 Decay&$e^{\pm}\mu^{\mp}$& $\mu^{+}\pi^{-}$ &$\mu^{-}\pi^{+}$ &$e^{+}\pi^{-}$ &$e^{-}\pi^{+}$ & $e^{\pm}\mu^{\mp}$& $\mu^{+}\pi^{-}$ & $\mu^{-}\pi^{+}$& $e^{+}\pi^{-}$ & $e^{-}\pi^{+}$  \\
\hline
Generation & \multicolumn{5}{|c|}{2958.0} & \multicolumn{5}{|c|}{$6.7\times10^{8}$} \\
 \hline
Channel                             &152.1&11.1&7.8 &11.8&8.6 &$3.6\times10^{7}$ &$5.2\times10^{7}$&$5.2\times10^{7}$&$5.3\times10^{7}$ &$5.2\times10^{7}$\\
\hline
$p_{T}(l,chg)>$ 1.0 GeV                 &18.7 &0.07&0.07&0.03&0.02&$5.4\times10^{6}$ &$6.2\times10^{6}$ &$6.2\times10^{6}$ & $6.3\times10^{6}$&$6.6\times10^{6}$\\
 \hline
 $ 2.0 \leq \eta_{chg,l} \leq 4.5$ &2.4  &0.0 &0.0 &0.0 &0.0 &$2.5\times10^{5}$ &$2.5\times10^{5}$ &$2.6\times10^{5}$ &$2.7\times10^{5}$ &$2.5\times10^{5}$\\
  \hline
$p_{T}(l + chg) >$ 2.0 GeV                 &0.9  &0.0 &0.0 &0.0 &0.0 &$8530.0$ &11272.0 &10358.0 &8225.0 &12795.0\\
 \hline
HERSCHEL                            &0.7  &0.0 &0.0 &0.0 &0.0 &$8530.0$ &11272.0 &10358.0 &8225.0 &12795.0\\
\hline
Sum   &1.4  & \multicolumn{4}{|c|}{0.0}  &17060.0 &\multicolumn{4}{|c|}{42650.0}\\
\hline
\end{tabular}
\caption{Predictions for the  ditau production at forward rapidities in $PbPb$ collisions at $\sqrt{s} = 5.5$ TeV  in exclusive and diffractive interactions considering the different decay channels.}
\label{table:sigma_tau}
\end{table}
\end{center}



Let's summarize our main results and conclusions. In this letter, we have investigated, for the first time, the dilepton production in diffractive and exclusive production at forward rapidities considering ultraperipheral $PbPb$ collisions at the LHC. We have derived predictions for the $e^+ e^-$, $\mu^+ \mu^-$ and $\tau^+ \tau^-$ cross sections taking into account of realistic cuts that can be implemented by the LHCb experiment in a future data analysis. Our results indicate that the background associated to the diffractive production can be strongly suppressed and the exclusive processes can be cleanly observed. We also demonstrated that selecting events with large acoplanarity, the diffractive events can be studied, allowing to constrain the modeling of the soft survival factor. For the $\tau^+ \tau^-$ production, the semi and purely leptonic decay channels were considered and the final results indicate that the study of the exclusive production can be performed. Finally, the predictions presented in this indicate that a future experimental analysis of the dilepton production at the LHCb is feasible and can be useful to  search for BSM physics.

\section*{Acknowledgements}
This work was  partially financed by the Brazilian funding
agencies CNPq, CAPES,  FAPERGS, FAPERJ and INCT-FNA (processes number 
464898/2014-5 and 88887.461636/2019-00).



\end{document}